\documentclass[twocolumn,showpacs,preprintnumbers,amsmath,amssymb,nofootinbib]{revtex4}

\usepackage{graphicx}
\usepackage{dcolumn}
\usepackage{bm}
\usepackage{amssymb}
\usepackage{amsmath}
\usepackage{amsthm}

\begin{document}


\title{Improved separation of soft and hard components in 
multiple Coulomb scattering}

\author{M.V.~Bondarenco}
 \email{bon@kipt.kharkov.ua}
 \affiliation{NSC Kharkov Institute of Physics and Technology, 1 Academic Street,
61108 Kharkov, Ukraine}

\date{\today}

\begin{abstract}
Evaluation of the angular distribution function of particles
scattered in an amorphous medium is improved by deforming the
integration path in the Fourier integral representation into the
complex plane. That allows us to present the distribution function
as a sum of two positive components, soft and hard, the soft
component being close to a Gaussian, and the hard component
vanishing in the forward direction, while including the Rutherford
asymptotics and all the power corrections to it at large scattering
angles. Detailed properties of those components, and their interplay
at intermediate deflection angles are discussed. Comparison with the
Moli\`{e}re theory is given.


\end{abstract}





\maketitle

\section{Introduction}

At passage of ultrarelativistic charged particles through amorphous
matter, they undergo multiple, essentially uncorrelated scattering
on atoms, typically through small angles. If the target is not too
thick, the longitudinal momentum of a high-energy particle may be
regarded as conserved. Then, the transport equation depends only on
the particle deflection angles, and is exactly solvable by means of
Fourier transformation \cite{Bothe-Wentzel}. However, conditions of
multiple Coulomb scattering on screened atomic nuclei may require
one to separately treat hard and soft contributions to the
distribution function, as was first pointed out by Williams
\cite{Williams}. That separation was cast in the form of a
large-thickness expansion by Moli\`{e}re
\cite{MoliereFirst,Moliere}, subsequently reviewed by Bethe
\cite{Bethe}, and nowadays is recognized as a standard procedure
(see \cite{Scott,Mott-Massey,Uch-Zol}). In modern practice, yet a
simplified approach is applied at times, retaining only the Gaussian
component with the root mean square angle inferred from Gaussian
fits \cite{Highland-Lynch-Dahl}, or derived analytically from the
Mol\`{e}re theory \cite{Bond-Shul}. But in high-statistics
experiments, non-Gaussian ``wings" are noticeable even for rather
thick targets.

Although Moli\`{e}re's expansion provides a formal background for
the theory, from the physical point of view it is not completely
satisfactory. It is known that, in principle, it does not converge
(see, e.g., \cite{Bielajew}), and yet, is comprised of oscillatory
functions of deflection angles, which do not admit independent
probabilistic interpretation. At the same time, there were recent
phenomenological indications that beyond the central Gaussian
region, the distribution function does not immediately switch to the
asymptotic power law  corresponding to single scattering, but
exhibits some transient behavior over a sizable range of angles
\cite{Taratin,Arleo}.

In case such a transition region does exist, the best option to
compute within it the distribution function would be to resum all
the non-Gaussian (at least, power-law) contributions through all
orders, as is done in other physical problems (see, e.g.,
\cite{Bondarenco-paper}). That typically leads to integral
representations for resummed quantities. But in the present case,
one can employ to this end a method \cite{Bethe,Scott}, in which the
original Fourier integral representation for the distribution
function is extended into the complex plane, and two principally
different (vertical and horizontal) parts of the integration path
are distinguished. That method so far has never developed to a
procedure superior to Moli\`{e}re's expansion; nonetheless, with
some improvements, it can be raised to that status, and provide a
different view on behavior of the angular distribution beyond the
central region. The key notion here is that integrals over the
mentioned parts of the path appear to be positive, and therefore may
be interpreted as hard and soft scattering components, coexisting at
any scattering angle. Comparison of those components will allow us
to determine the width of the transition region between Gaussian and
Rutherford regions in the aggregate distribution, and assess the
significance of resummation of all the plural hard-scattering
contributions.


\section{Preliminary considerations}

\subsection{Fourier-Bessel solution of the transport equation}

The probability distribution of fast particles scattered through
small angles $\theta$ in an amorphous medium,
$f(\theta,l)=\frac{dw}{d^2\theta}$, is governed by the transport
equation
\begin{equation}\label{transport-eq}
\frac{\partial f}{\partial l}= n\int
d\sigma(\chi)\left[f(\bm\theta-\bm\chi,l)-f(\theta,l)\right],
\end{equation}
where $d\sigma(\chi)=d^2\chi\frac{d\sigma}{d^2\chi}$ is the
differential cross-section of particle scattering on one atom
through angle $\chi$, $n$ is the density of atoms in the medium, and
$l$ the traversed target thickness. Equation~(\ref{transport-eq})
conserves the normalization:
\begin{equation}\label{norm=1}
\int d^2\theta f(\theta,l)=1,
\end{equation}
Solution of Eq.~(\ref{transport-eq}) satisfying the initial
condition $f(\theta,0)=\delta(\bm\theta)$ is obtained by means of
Fourier-Bessel transformation:
\begin{subequations}\label{2DFourier}
\begin{eqnarray}
f(\theta,l)&=&\int \frac{d^2\rho}{(2\pi)^2}
e^{i\bm{\rho}\cdot\bm{\theta}-nl\int d\sigma(\chi)
(1-e^{-i\bm{\rho}\cdot\bm{\chi}})}\label{}\\
&\equiv&\frac{1}{2\pi}\int_0^{\infty} d\rho\rho J_0(\rho\theta)
e^{-nl\int d\sigma(\chi) \left[1-J_0(\rho\chi)\right]}.\label{3b}
\end{eqnarray}
\end{subequations}

In some applications, one may be concerned rather with the projected
angle distribution, which is given by a 1-dimensional Fourier
transformation:
\begin{eqnarray}\label{f-proj-def}
f(\theta_x,l)&=&\int_{-\infty}^{\infty}
d\theta_yf(\theta,l)\nonumber\\
&=&\int_{-\infty}^{\infty} \frac{d\xi}{2\pi} e^{i\xi\theta_x-nl\int
d\sigma(\chi) \left[1-J_0(\xi\chi)\right]}.
\end{eqnarray}
We shall denote distribution functions (\ref{2DFourier}) and
(\ref{f-proj-def}) by the same letter $f$, distinguishing them just
by notation of their angle arguments.

\subsection{Thick targets: Moli\`{e}re's theory}

At significant target thickness, the random walk in the plane of
deflection angles (which may be viewed as transverse vectors) must
reduce to diffusion. In generic integral representations
(\ref{2DFourier}) and (\ref{f-proj-def}), that comes about as
follows: at large $nl$, the exponential in their integrands is
rapidly decreasing, therefore the contributing $\rho$ or $\xi$ are
small, permitting one to expand the exponent to leading order in
their values. However, naive expansion
$1-J_0(\rho\chi)\simeq{\rho^2\chi^2}/{4}$ in the integrand gives a
logarithmically diverging variance $\int d\sigma(\chi)\chi^2$, given
that the physical differential cross-section of fast charged
particle scattering on one atom through large angles obeys the
Rutherford asymptotics
\begin{equation}\label{Ruth-dsigma}
\frac{d\sigma}{d\chi}\underset{\chi/\chi'_a\to\infty}\simeq
\frac{8\pi Z^2\alpha^2}{p^2\chi^3},
\end{equation}
with $p$ being the particle momentum, $Z$ the nucleus charge, and
$\alpha$ the fine structure constant. A more accurate calculation
\cite{Bethe} shows that the small-$\rho$ asymptotics of the exponent
in (\ref{2DFourier}), (\ref{f-proj-def}) involves a factor
logarithmically depending on $\rho$:
\begin{equation}\label{J0-square-log}
nl\int d\sigma(\chi)
\left[1-J_0(\rho\chi)\right]\underset{\rho\chi'_a\to0}\simeq\frac{\chi_c^2\rho^2}2\ln\frac{2}{\chi'_a\rho},
\end{equation}
and thereby spoiling the Gaussianity of the Fourier-Bessel integral.
Here $\chi_c^2(l)=4\pi nl Z^2\alpha^2/p^2$, and the screening angle
$\chi'_a\sim1/R_{a}p$, with $R_a$ being the atomic radius,
characterizes the scale of angles at which the singularity in
(\ref{Ruth-dsigma}) is tamed.\footnote{In terms of the exact
scattering differential cross-section
$\frac{d\sigma}{d\chi}=\frac{8\pi Z^2\alpha^2}{p^2\chi^3}q(\chi)$,
with $\chi^{-4}q(\chi)\underset{\chi/\chi'_a\to0}\to\text{const}>0$
and $q(\chi)\underset{\chi/\chi'_a\to\infty}\to1$, the screening
angle expresses as
\begin{equation}\label{dsigma-J0->ln}
\ln\chi'_a=\int dq(\chi)\ln\chi+\gamma_{\text{E}}-1,
\end{equation}
where $\gamma_{\text{E}}$ is the Euler's constant. This definition
\cite{Bethe} differs from the more conventional $\chi_a$
\cite{Moliere} by terms $\gamma_{\text{E}}-1/2=0.077$, but
numerically, the difference is small. With definition
(\ref{dsigma-J0->ln}), the right-hand side (rhs) of
Eq.~(\ref{J0-square-log}) is the shortest, facilitating the
following calculations. Note, too, that while
Eq.~(\ref{J0-square-log}) was written for pure elastic scattering,
inelastic contributions can also be incorporated there \cite{Fano},
just by redefining $\chi'_a$ and $\chi_c$.}
Thus, the diffusion here is anomalous, but only marginally, in the
sense that the anomaly is logarithmic instead of a power law. That
implies that the distribution function does \emph{not} approach a
L\'{e}vy distribution \cite{Uch-Zol}, albeit is not strictly
Gaussian either.

Ratio ${\chi_c^2}/{\chi'^2_a}$ essentially measures the target thickness in units of the radiation length $X_0$:
\[
\frac{\chi_c^2}{\chi'^2_a}=\frac{\pi}{\alpha
\gamma^2\chi'^2_a\ln\frac{\text{const}}{2\gamma\chi'_a}}\frac{l}{X_0},
\]
with $\text{const}\sim1$, and $\gamma\chi'_a$ expressible in terms
of $X_0$, as well [see, e.g., \cite{Bond-Shul}, Eq.~(42)]. For
instance, ratio ${\chi_c}/{\chi'_a}=10^2$ corresponds to targets of
solid materials of a few millimeter thickness. In what follows, we
will measure the target thickness in $Z$-independent fashion, merely
in units of ${\chi_c^2}/{\chi'^2_a}$.

Approximation (\ref{J0-square-log}) appreciably simplifies the
structure of integrals (\ref{2DFourier}), (\ref{f-proj-def}), but
their evaluation still involves non-trivial aspects. Intuitively, it
is clear that the diffusion, at least at typical angles, must be
close to Gaussian, although with possible logarithmic deviations. To
tackle those, Moli\`{e}re \cite{Moliere} assumed that the typical
deflection angle is $\chi_c\sqrt{B}$, with $B$ such that the
difference of logarithmically large parameters $B-\ln
B-\ln\frac{\chi_c^2}{\chi'^2_a}$ is a constant of the order of unity
(conventionally set to be zero). Therewith, $B(\chi_c^2/\chi'^2_a)$
is a Lambert (or product logarithm) function, asymptotically equal
$B\underset{\chi_c\gg\chi'_a}\simeq\ln\left(\frac{\chi_c^2}{\chi'^2_a}\ln\frac{\chi_c^2}{\chi'^2_a}\right)$,
and the rhs of (\ref{J0-square-log}) rewrites as
\[
\frac{\chi_c^2\rho^2}2\ln\frac{2}{\chi'_a\rho}=\frac{u^2}{4}-\frac{u^2}{4B}\ln\frac{u^2}{4},
\]
where $u=\chi_c\sqrt{B}\rho$. As long as the logarithmic dependence
on the rescaled integration variable $u$ in the exponent appears to
be inversely proportional to the large parameter $B$, that suggests
expanding this part of the exponential into power series and
formally integrating termwise:
\begin{equation}\label{Moliere-expansion}
f(\theta,l)=\frac1{2\pi\chi_c^2
B}\sum_{k=0}^{\infty}\frac1{B^k}f^{(k)}\left(\frac{\theta}{\chi_c\sqrt{B}}\right),
\end{equation}
with
\begin{equation}\label{fk-Moliere-def}
f^{(k)}(\Theta)=\frac1{k!}\int_0^{\infty}duuJ_0(\Theta
u)e^{-u^2/4}\left(\frac{u^2}4\ln\frac{u^2}4\right)^k.
\end{equation}
Note that the expansion parameter $B^{-1}$ here is only
logarithmically small, but for $B\geq 4.5$, i.e., $\chi_c\gg
10\chi'_a$, expansion (\ref{Moliere-expansion}) is reported to work reasonably well \cite{Moliere,Bethe}. An important
consequence of (\ref{fk-Moliere-def}) is that for all $k\geq1$,
\begin{equation}\label{int-fk-Moliere=0}
\int d^2\theta f^{(k)}(\theta)\equiv0.
\end{equation}
Hence, functions $f^{(k)}$ at $k\geq1$ are not everywhere positive,
and do not admit probabilistic interpretation.

Analyzing integrals (\ref{fk-Moliere-def}), one finds that at large
$\Theta$, components of (\ref{Moliere-expansion}) behave as
$f^{(0)}(\Theta)=2e^{-\Theta^2}$, which corresponds to a perfect
Gaussian, and $f^{(1)}(\Theta)\sim \Theta^{-4}$, which reflects the
Rutherford asymptotics $f(\theta)\simeq
\frac{nl}{2\pi\theta}\frac{d\sigma}{d\theta}$. For $k\geq2$,
$f^{(k)}(\Theta)\sim \Theta^{-2-2k}$ times logarithmic factors
(which will be determined below). Further analysis reveals that, in
fact, functions $f^{(k)}$ for $k\geq1$ make several
oscillations,\footnote{That owes to the fact that as $k$ increases,
factor $e^{-u^2/4}\left(\frac{u^2}4\ln\frac{u^2}4\right)^k$ in the
integrand of (\ref{fk-Moliere-def}) becomes sharply peaking at
$u\sim2\sqrt{k}$. Therewith, at fixed $\Theta$ and increasing $k$,
integral (\ref{fk-Moliere-def}) tends to
\begin{equation}\label{ln^kk}
f^{(k)}(\Theta)\sim2\ln^k k J_0(2\sqrt{k}\Theta).
\end{equation}
For $\Theta=0$, the latter scaling law was quoted in
\cite{Bielajew}. } which are much stronger than the asymptotic
power-law ``tails". At moderate $\chi_c/\chi'_a$, they may cause a
spurious warp in between the Gaussian and Rutherford regions. Yet,
despite the factor $k!$ in the denominator in the rhs of
(\ref{fk-Moliere-def}), functions $f^{(k)}$ \emph{grow} with $k$
faster than exponentially [see Eq.~(\ref{ln^kk})]. Therefore, in
principle, series (\ref{Moliere-expansion}) diverges, though it may
still serve as an asymptotic expansion in the limit $nl\to\infty$.


\subsection{Thin targets: Power and logarithmic corrections to the Rutherford asymptotics}\label{subsec:Glauber-expansion}

Even though at typical angles the number of scatterings in any
macroscopic target is very large, at significant deflection angles
the distribution function may be determined by just a few hard
scatterings. It can thus be useful to expand the distribution
function into perturbation series
\begin{equation}\label{series-f-thetax}
f(\theta_x,l)=\sum_{k=1}^{\infty}(nl)^kf_k(\theta_x),
\end{equation}
and study the behavior of its components $f_k(\theta_x)$ at large
$\theta_x$.

The lowest-order terms of (\ref{series-f-thetax}) are
\begin{eqnarray}\label{f1x-Ruth-asympt}
f_1(\theta_x)&=&\frac{1}{2\pi}\int_{-\infty}^{\infty}
d\xi\cos(\xi\theta_x) \int d\sigma(\chi)
\left[J_0(x\chi)-1\right]\nonumber\\
&\equiv&\frac{1}{2\pi}\int_{-\infty}^{\infty} d\xi e^{i\xi\theta_x}
\int_{-\infty}^{\infty} d\chi_x\frac{d\sigma}{d\chi_x}
\left(e^{-ix\chi_x}-1\right)\nonumber\\
&=&\frac{d\sigma}{d\theta_x}-\sigma\delta(\theta_x)\underset{\theta_x/\chi'_a\to\infty}\sim
\frac{\chi_c^2}{2nl\theta_x^3},
\end{eqnarray}
and
\begin{eqnarray}
f_2(\theta_x)&=&\frac{1}{4\pi}\int_{-\infty}^{\infty} d\xi
e^{i\xi\theta_x} \left[\int_{-\infty}^{\infty}
d\chi_x\frac{d\sigma}{d\chi_x}
\left(e^{-i\xi\chi_x}-1\right)\right]^2\nonumber\\
&=&\frac12\int_{-\infty}^{\infty} d\chi_x\frac{d\sigma}{d\chi_x}\frac{d\sigma}{d(\theta_x-\chi_x)}-\sigma\frac{d\sigma}{d\theta_x}+\frac{\sigma^2}2\delta(\theta_x).\nonumber\\
\label{proj-f2}
\end{eqnarray}
The dominant contribution to the integral term in (\ref{proj-f2})
comes from neighborhoods of two points: $\chi_x=0$, where
$\frac{d\sigma}{d(\theta_x-\chi_x)}$ may be approximated by a
constant, and $\chi_x=\theta_x$, where
$\frac{d\sigma}{d\chi_x}\simeq \frac{d\sigma}{d\theta_x}$. The
corresponding asymptotics of the integral thus equals
$\frac12\int_{-\infty}^{\infty}
d\chi_x\frac{d\sigma}{d\chi_x}\frac{d\sigma}{d(\theta_x-\chi_x)}\simeq
\sigma\frac{d\sigma}{d\theta_x}$, but it is exactly canceled by the
second term of (\ref{proj-f2}). Therefore, to determine the
asymptotics of $f_2$, one has to expand the slowly varying factors
in the integrand to higher orders:
\begin{eqnarray}
f_2(\theta_x)\!&\underset{\theta_x/\chi'_a\to\infty}\simeq&\!\int d\chi_x\frac{d\sigma}{d\chi_x}\!\left(\!-\chi_x\frac{d}{d\theta_x}\frac{d\sigma}{d\theta_x}+\frac{\chi_x^2}2\frac{d^2}{d\theta_x^2}\frac{d\sigma}{d\theta_x}\right)\nonumber\\
&=&\frac{1}2\frac{d^2}{d\theta_x^2}\frac{d\sigma}{d\theta_x}\int
d\chi_x\chi_x^2\frac{d\sigma}{d\chi_x}.\label{}
\end{eqnarray}
Here
$\frac{d^2}{d\theta_x^2}\frac{d\sigma}{d\theta_x}\simeq\frac{6\chi_c^2}{nl\theta_x^5}$,
and $\int_{\sim-\theta_x}^{\sim \theta_x}
d\chi_x\chi_x^2\frac{d\sigma}{d\chi_x}\simeq
\frac{\chi_c^2}{nl}\ln\frac{\theta_x}{\chi'_a}$, wherewith
\begin{equation}\label{}
(nl)^2f_2(\theta_x)\underset{\theta_x/\chi'_a\to\infty}\simeq\frac{3\chi_c^4}{\theta_x^5}\ln\frac{\theta_x}{\chi'_a}.
\end{equation}
Hence, if one considers a ``form factor" $\theta_x^3 f(\theta_x)$,
which vanishes at $\theta_x=0$, and tends to a constant as
$\theta_x/\chi'_a\to\infty$, it appears to be a nonmonotonous
function of $\theta_x$, and overshoots the latter constant at some
intermediate $\theta_x$. That salient feature of the multiple
Coulomb scattering angular distribution was confirmed experimentally
(see \cite{Hanson,Bethe}).

Similarly, it can be proven that higher-order terms
in (\ref{series-f-thetax}) are all positive and asymptotically
scale as
\begin{equation}\label{fk-as-x}
(nl)^k
f_k(\theta_x)\underset{\theta_x/\chi'_a\to\infty}\simeq\frac{k(2k-1)!!\chi_c^{2k}}{2\theta_x^{1+2k}}\ln^{k-1}\frac{\theta_x}{\chi'_a}.
\end{equation}

For polar angle distribution
\begin{equation}\label{series-f-theta}
f(\theta,l)=\sum_{k=1}^{\infty}(nl)^k f_k(\theta),
\end{equation}
the asymptotics of the leading terms of the expansion is
\begin{equation}\label{f1-Ruth}
f_1(\theta)=\frac{d\sigma}{d^2\theta}+\sigma\delta(\bm\theta)\underset{\theta/\chi'_a\to\infty}\simeq\frac{\chi_c^2}{\pi
nl\theta^4},
\end{equation}
\begin{equation}\label{f2-asympt}
f_2(\theta)\underset{\theta/\chi'_a\to\infty}\simeq\frac14\triangle_{\theta}\frac{d\sigma}{d^2\theta}\int_0^{\sim\theta/2}
\!d\chi\chi^2\frac{d\sigma}{d\chi}\simeq\frac{8\chi_c^4}{\pi(nl)^2\theta^6}\!\ln\frac{\theta}{\chi'_a},
\end{equation}
and generally
\begin{equation}\label{fk-as}
(nl)^k f_k(\theta)\underset{\theta/\chi'_a\to\infty}\simeq
\frac{kk!2^{k-1}\chi_c^{2k}}{\pi\theta^{2+2k}}\ln^{k-1}\frac{\theta}{\chi'_a}.
\end{equation}

Note that the coefficients at logarithms in Eqs.~(\ref{fk-as-x}),
(\ref{fk-as}) turn out to be sizable already at $k=1$, and grow with
$k$ factorially. Thus, at moderately large $\theta$, it would be
advantageous to sum such contributions through all orders.
Resummations of that kind are usually carried out via Borel
transformation \cite{Borel-summ}. But in our case, construction of a
new integral representation is unnecessary, as long as the original
integral representation (\ref{2DFourier}) or (\ref{f-proj-def}) is
already well suited for that purpose. Below we will derive
corresponding resumming expressions directly from integrals
(\ref{2DFourier}) and (\ref{f-proj-def}).

\section{Analysis in the complex plane}

Since we are interested in the case when the number of collisions is
high, the exponent in integrals (\ref{2DFourier}),
(\ref{f-proj-def}) will generally assume large values. The modern
approach to deriving asymptotics of such integrals consists in
extending the integral into a complex plane. With an appropriate
choice of the integration path, the integrand can be made
non-oscillatory, which substantially alleviates derivation of the
asymptotics of the integral. In application to multiple Coulomb
scattering distributions, such a deformation procedure was first
suggested by Bethe (see Appendix A in \cite{Bethe}, and also
\cite{Scott}), but served mainly for the purpose of deriving the
coefficients of large-angle power asymptotic terms \cite{Scott}, or
combining just a few such terms to an expression, which still worked
only in a limited domain of $\theta$ (at large $\theta$)
\cite{Bethe}. Here we are going to handle the entire sequence of
asymptotic terms simultaneously, but in order to make it applicable
\emph{everywhere}, the definition of the integration path must be
improved. The path extension problem appears to be technically
simpler for the projected angle distribution, which was not
considered in \cite{Bethe} at all, and which we consider here first.

\subsection{Projected angle distribution}\label{subsec:proj-angle}

\begin{figure}
\includegraphics{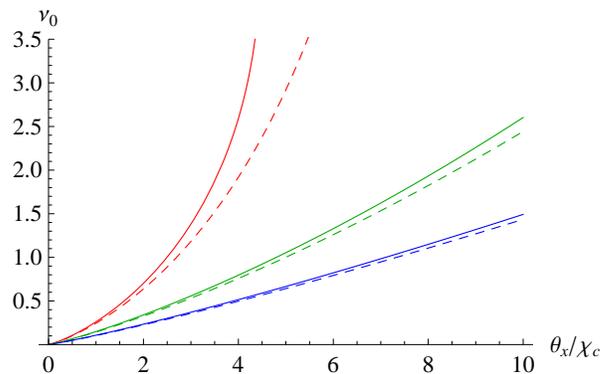}
\caption{\label{fig:lambda0-proj-angle} Solid curves, behavior of
solution of the corner point equation for the polar angle
distribution [Eq.~(\ref{approx-saddle-point-eq-proj})]. Dashed curves, approximation (\ref{lambda0-x-log-approx}).
Red, for $\chi_c/\chi'_a=10$; green, for $\chi_c/\chi'_a=10^2$;
blue, for $\chi_c/\chi'_a=10^3$.}
\end{figure}

The diffusion approximation to Eq.~(\ref{f-proj-def}) reads (see
footnote 1)
\begin{equation}\label{1DFourier}
f(\theta_x,l)\underset{\chi_c/\chi'_a\to\infty}\simeq\frac1{\pi\chi_c}\mathfrak{Re}\int_0^{\sim\chi_c/\chi'_a}
d\kappa e^{i\frac{\theta_x}{\chi_c}\kappa+\frac{\kappa^2}2
\ln\frac{\chi'_a\kappa}{2\chi_c}},
\end{equation}
where we set $\kappa=\xi\chi_c$. When extending this integral to the
plane of complex $\kappa$, it is found that its integrand has a
single saddle point obeying the equation
\begin{equation}\label{saddle-point-eq-x}
\frac{\partial}{\partial\kappa}\!\!\left(\!i\frac{\theta_x}{\chi_c}\kappa+\frac{\kappa^2}2 \ln\frac{\chi'_a\kappa}{2\chi_c}\right)\!\!\Bigg|_{\kappa=\kappa_0}
\!\!=i\frac{\theta_x}{\chi_c}+\kappa_0\!\left(\!\ln\frac{\kappa_0\chi'_a}{2\chi_c}+\frac12\right)\!=0.
\end{equation}
As long as Eq.~(\ref{saddle-point-eq-x}) is transcendental, only its
approximate solution can be expressed explicitly, which, though,
will suit us at the present stage. We can choose an approximation to
the solution of (\ref{saddle-point-eq-x}), which is strictly
imaginary:
\begin{equation}\label{lambda0-x-log-approx}
\kappa_0=i\nu_0, \qquad
\nu_0\approx\frac{\theta_x}{\chi_c\ln\left(\frac{2\chi_c^2}{\chi'_a\theta_x}\ln\frac{2\chi_c^2}{\chi'_a\theta_x}\right)},
\end{equation}
with a proviso that this formula is good only for
$\chi_c/\chi'_a\gg10$ (and $\theta_x<2\chi_c^2/\chi'_a$, which is
usually fulfilled in practice). To illustrate the accuracy of
approximation (\ref{lambda0-x-log-approx}), in
Fig.~\ref{fig:lambda0-proj-angle} it is plotted along with the exact
solution of equation
\begin{equation}\label{approx-saddle-point-eq-proj}
\frac{\theta_x}{\chi_c}+\nu_0\left(\ln\frac{\chi'_a\nu_0}{2\chi_c}+\frac12\right)=0,
\end{equation}
obtained from (\ref{saddle-point-eq-x}) by neglecting $\ln i$. It
clearly indicates that approximation (\ref{lambda0-x-log-approx})
begins to fail for $\chi_c/\chi'_a\sim10$.

\begin{figure}
\includegraphics{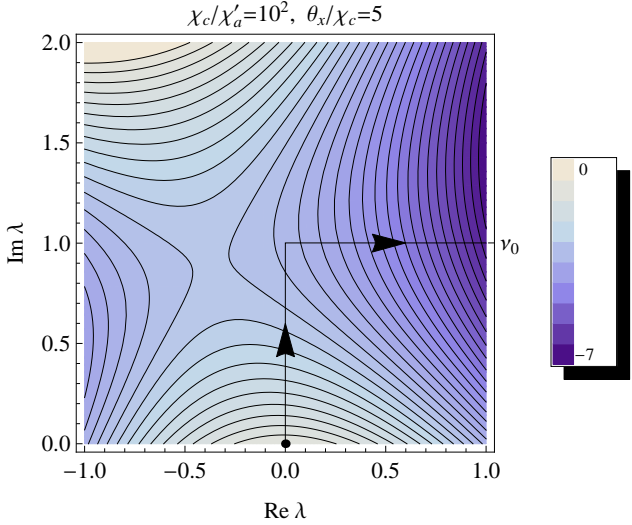}
\caption{\label{fig:path-proj-angle} Gradient plot of function
$\mathfrak{Re}\left(i\frac{\theta_x}{\chi_c}\kappa+\frac{\kappa^2}2
\ln\frac{\chi'_a\kappa}{2\chi_c}\right)$ [the real part of the
exponent in Eq.~(\ref{1DFourier})] in the upper half-plane of
complex integration variable $\kappa$, for exemplary values of
$\chi_c$ and $\theta_x$. The deformed integration path is drawn
by the black line, with $\nu_0$ evaluated by
Eq.~(\ref{lambda0-x-log-approx}).
}
\end{figure}

The logarithmic factor in the exponent in (\ref{1DFourier}) induces
a singularity of the integrand at the origin, coinciding with the
lower endpoint of the integration interval. The steepest descent
path must then start at the origin, and go toward the saddle point.
For simplicity of the resulting integral, though, we direct it
strictly along the imaginary axis, rewriting the integration
variable as $\kappa=i\nu$. After reaching a point $\kappa_0$ defined
by Eq.~(\ref{approx-saddle-point-eq-proj}), the path must turn to
the right and proceed along the steepest descent path, but again,
for simplicity, we just direct it parallel to the real axis (see
Fig.~\ref{fig:path-proj-angle}). Ultimately, the distribution
function splits to a sum of two real-variable integrals:
\begin{equation}\label{fthetax=fhard+fsoft}
f(\theta_x,l)=f_{h}(\theta_x,l)+f_{s}(\theta_x,l),
\end{equation}
where
\begin{equation}\label{fx-hard-def}
f_{h}(\theta_x,l)=\frac1{\pi\chi_c}\int_0^{\nu_0(\theta_x)} d\nu
e^{-\frac{\theta_x}{\chi_c}\nu+\frac{\nu^2}2
\ln\frac{2\chi_c}{\chi'_a\nu}}\sin\frac{\pi\nu^2}{4},
\end{equation}
\begin{equation}\label{fx-soft-def}
f_{s}(\theta_x,l)=\frac1{\pi\chi_c}\mathfrak{Re}\int_{i\nu_0(\theta_x)}^{\sim\chi_c/\chi'_a}
d\kappa e^{i\frac{\theta_x}{\chi_c}\kappa+\frac{\kappa^2}2
\ln\frac{\chi'_a\kappa}{2\chi_c}}.
\end{equation}

Below we will show that in spite of the admitted simplification of
the integration path, integrals (\ref{fx-hard-def}),
(\ref{fx-soft-def}) can be robustly interpreted as hard and soft
scattering components. Our task now is to investigate their
properties.

\paragraph{Hard component}


Component $f_{h}$ proves to be positive everywhere, even for an
approximate solution of the saddle-point equation, insofar as
typical contributing $\nu$ in Eq.~(\ref{fx-hard-def}) are always
$\lesssim1$, entailing $\sin\frac{\pi\nu^2}4>0$. Furthermore, almost
everywhere it is tolerable to replace in (\ref{fx-hard-def})
$\sin\frac{\pi\nu^2}4\approx\frac{\pi\nu^2}4$. That is strictly
justified in limits of either large or small $\theta_x/\chi_c$: If
$\theta_x/\chi_c\ll1$, that becomes possible because the upper
integration limit tends to zero, leaving
\begin{eqnarray}\label{fhard-x-lowthetax}
f_{h}(\theta_x,l)&\underset{\theta_x/\chi_c\ll1}\simeq&
\frac1{4\chi_c}\int_0^{\frac{\theta_x}{\chi_c\ln\left(\frac{2\chi_c^2}{\chi'_a\theta_x}\ln\frac{2\chi_c^2}{\chi'_a\theta_x}\right)}}d\nu\nu^2\nonumber\\
&=&\frac{\theta_x^3}{12\chi_c^4\ln^3\left(\frac{2\chi_c^2}{\chi'_a\theta_x}\ln\frac{2\chi_c^2}{\chi'_a\theta_x}\right)}.
\end{eqnarray}
If $\theta_x/\chi_c\to\infty$, the sine in (\ref{fx-hard-def}) can
be linearized by virtue of the rapid decrease of factor
$e^{-\frac{\theta_x}{\chi_c}\nu}$ in the integrand. Therewith,
expansion of the rest of the exponential into Maclaurin series
yields the Rutherford law (\ref{f1x-Ruth-asympt}), along with power
corrections to it (beyond the leading logarithmic accuracy):
\begin{eqnarray}
f_{h}(\theta_x,l)&\underset{\theta_x/\chi_c\to\infty}\simeq&\frac1{4\chi_c}\int_0^{\infty}\! d\nu\nu^2 e^{-\frac{\theta_x}{\chi_c}\nu}\!\left(\!1+\frac{\nu^2}2 \ln\frac{2\chi_c}{\chi'_a\nu}\right)\nonumber\\
&=&\frac{\chi_c^2}{2\theta_x^3}+3\frac{\chi_c^4}{\theta_x^5}\left[\ln\frac{2\theta_x}{\chi'_a}-\psi(5)\right],\label{Ruth-x+corr}
\end{eqnarray}
with $\psi(z)=\Gamma'(z)/\Gamma(z)$ being the digamma function.
Clearly, integral (\ref{fx-hard-def}) resums also all the higher
power corrections to the Rutherford asymptotics.

\begin{figure}
\includegraphics{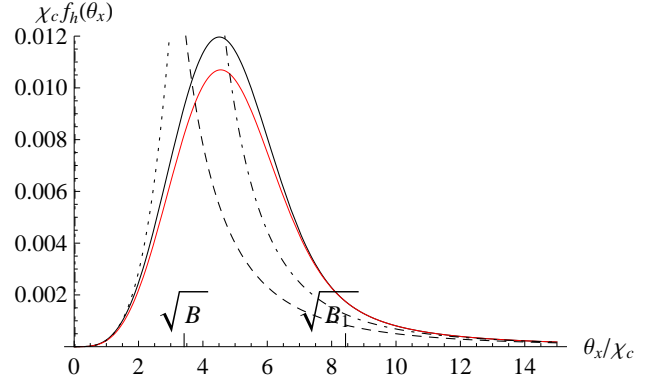}
\caption{\label{fig:f-x-hard} Hard component of the projected angle
distribution function at $\chi_c/\chi'_a=10^2$, built by
Eqs.~(\ref{fx-hard-def}), (\ref{approx-saddle-point-eq-proj}) (solid black
curve), and by Eqs.~(\ref{fx-hard-def}), (\ref{lambda0-x-log-approx}) (solid red curve). Dashed curve, Rutherford asymptotics
(\ref{f1x-Ruth-asympt}). Dot-dashed, Rutherford asymptotics with the
first power correction, Eq.~(\ref{Ruth-x+corr}). Dotted,
low-$\theta_x$ asymptotics (\ref{fhard-x-lowthetax}).}
\end{figure}

The fact that the component $f_{h}(\theta_x)$ vanishes in both
extremes $\theta_x/\chi_c\to0$ and $\theta_x/\chi_c\to\infty$
implies that it must peak somewhere in between [see
Fig.~\ref{fig:f-x-hard}]. From the analysis of integral
(\ref{fx-hard-def}), one generally concludes that the summit of
$f_{h}(\theta_x)$ must be reached when $\nu_0\sim\chi_c/\theta_x$,
i.e., $\theta_x\sim\chi_c\sqrt{B(\chi_c^2/\chi'^2_a)}$, which is
nothing but Moli\`{e}re's typical angle. More precisely, that
corresponds to the rising slope of the peak, while the maximum is
located at a somewhat greater $\theta_x$ (see
Fig.~\ref{fig:f-x-hard}). The end of the region where resummation
effects are strong may be assessed from equating the Rutherford
asymptotic term to the \textit{doubled} next-to-leading-order power
correction in (\ref{Ruth-x+corr}):
$\frac{\chi_c^2}{2\theta_x^3}=2\times3\frac{\chi_c^4}{\theta_x^5}\left[\ln\frac{2\theta_x}{\chi'_a}-\psi(5)\right]$,
i.e., $\theta_x=\chi_c\sqrt{B_1}$, where
$B_1=6B(24e^{-2\psi(5)}\chi_c^2/\chi'^2_a)$. Due to the sizable
numerical coefficients involved therein, interval
\[
\chi_c\sqrt{B}<\theta_x<\chi_c\sqrt{B_1}\qquad \text{(semihard
region)}
\]
appears to be even wider than the soft central region
$0<\theta_x<\chi_c\sqrt{B}$.

Besides that, it is noteworthy that $f_{h}(\theta_x)$ does not lie
between its asymptotes (in particular, it goes well above the
Rutherford asymptote). This (or rather the corresponding feature for
$f_{h}(\theta)$ proven in the next subsection) may be responsible
for the empirical controversies mentioned in the Introduction.

\paragraph{Soft component}

\begin{figure}
\includegraphics{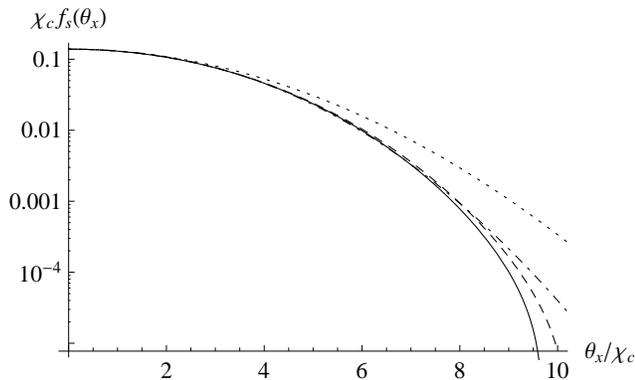}
\caption{\label{fig:f-x-soft} Soft component of the projected angle
distribution function at $\chi_c/\chi'_a=10^2$, built by
Eqs.~(\ref{fx-soft-def}) and (\ref{lambda0-x-log-approx}) (solid
curve). Dashed curve, the same evaluated for the corner point defined by (\ref{lambda0-x-log-approx}). Dot-dashed, the quasi-Gaussian approximation, Eqs.~(\ref{fsoft-x-Gauss}), (\ref{g0x}), with $C=2.2$. Dotted, Moli\`{e}re's $f^{(0)}$ for the projected angle
distribution.}
\end{figure}

Next, we inspect the soft component, which is defined by integral
(\ref{fx-soft-def}). This integral is close to Gaussian form, so its
fastest dependence on $\theta_x$ stems from the value of the
exponential at the endpoint:
\[
e^{-\frac{\theta_x}{\chi_c}\nu_0+\frac{\nu_0^2}2\ln\frac{2\chi_c}{i\chi'_a\nu_0}}\simeq
e^{-\frac{\theta_x}{2\chi_c}\nu_0},
\]
where we used the saddle point equation (\ref{saddle-point-eq-x})
within the accuracy to which we neglected $\ln i$ in
Eq.~(\ref{approx-saddle-point-eq-proj}). To account for the rest of
the $\theta_x$-dependence, the simplest way might be to replace in
the relation
\begin{equation}\label{f=expg}
f_{s}(\theta_x,l)=e^{-\frac{\theta_x}{2\chi_c}\nu_0(\theta_x,l)}g(\theta_x,l)
\end{equation}
the relatively slowly varying factor $g(\theta_x,l)$ by its value in
the origin,
\begin{equation}\label{g0x}
g(0,l)=f(0,l)=\frac1{\pi\chi_c}\int_0^{\sim\chi_c/\chi'_a} d\kappa
e^{-\frac{\kappa^2}2 \ln\frac{2\chi_c}{\chi'_a\kappa}}.
\end{equation}
More precisely, the width of $g$ is
$\theta_x\sim\chi_c\ln\frac{2\chi_c^2}{\chi'_a\theta_x}$, whereas
that of $f_s$ is
$\theta_x\sim\chi_c\sqrt{\ln\frac{2\chi_c^2}{\chi'_a\theta_x}}$,
which is narrower, but not by a very large factor. So, in practice
it would be certainly worth taking into account also the slope of
$g(\theta_x,l)$ in the origin. That can be implemented to the
structure of the leading exponential in Eq.~(\ref{f=expg}) by
approximating
\begin{equation}\label{gx-approx}
g(\theta_x,l)\to g(0,l)e^{-\frac{\theta_x^2\ln
C}{2\chi_c^2\ln^2\frac{2\chi_c^2}{\chi'_a\theta_x}}},
\end{equation}
with $C\approx2.2$. Combining (\ref{f=expg}),
(\ref{lambda0-x-log-approx}) and (\ref{gx-approx}), we obtain a
quasi-Gaussian structure
\begin{equation}\label{fsoft-x-Gauss}
f_{s}(\theta_x,l)\approx
f(0,l)e^{-\frac{\theta_x^2}{2\chi_c^2\ln\left[\frac{2\chi_c^2}{C\chi'_a\theta_x}\ln
\frac{2\chi_c^2}{\chi'_a\theta_x}\right]}}.
\end{equation}
It resembles the zeroth-order approximation
$f^{(0)}(\theta/\chi_c\sqrt{B})$ of Moli\`{e}re's expansion (applied
to the projected angle distribution), but has a more precise
normalization (\ref{g0x}), and yet involves $\theta_x$ under the
logarithm in the denominator of the exponent. Due to the latter
dependence, (\ref{fsoft-x-Gauss}) is narrower than Moli\`{e}re's
$f^{(0)}$ at $\theta_x>\chi_c$, i.e., in fact, at typical angles
(see Fig.~\ref{fig:f-x-soft}). A narrowing of that kind was
empirically found in \cite{Hanson}. Besides that, the integral of
(\ref{fsoft-x-Gauss}) over $\theta_x$, in contrast to the integral
of the zeroth component of Moli\`{e}re's expansion, is somewhat less
than unity, leaving a part of the probability for $f_{h}$.


\begin{figure}
\includegraphics{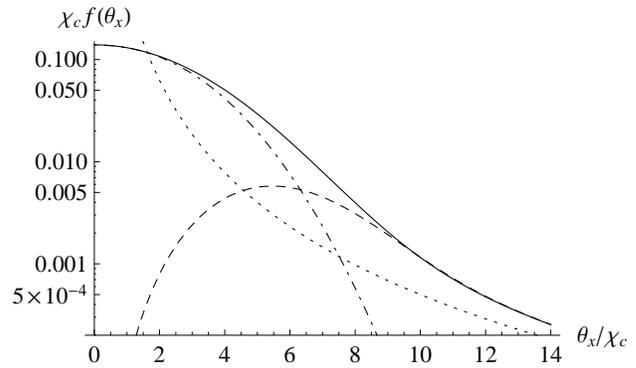}
\caption{\label{fig:proj-angle} Relative contributions of the hard
[dashed curve, Eqs.~(\ref{fx-hard-def}),
(\ref{lambda0-x-log-approx})] and soft [dot-dashed curve,
Eq.~(\ref{fsoft-x-Gauss})] components to the aggregate projected
angle distribution [solid curve, Eq.~(\ref{1DFourier})], for
$\chi_c/\chi'_a=10^2$. The sum of thus computed hard and soft
component is virtually indistinguishable from the solid curve. The
dotted curve shows the Rutherford asymptotics
(\ref{f1x-Ruth-asympt}).}
\end{figure}

\paragraph{Aggregate distribution}

The circumstance that components (\ref{fx-hard-def}),
(\ref{fx-soft-def}) in decomposition (\ref{fthetax=fhard+fsoft})
peak at different $\theta_x$ might potentially lead to appearance of
a secondary bump in the aggregate distribution. To check whether
this happens in reality, let us first assess the scale at which
$f_{s}(\theta_x)$ and $f_{h}(\theta_x)$ become commensurable. For
large $\chi_c/\chi'_a$, that occurs at relatively large $\theta_x$,
allowing one, oversimplistically, to employ the Rutherford
asymptotics for $f_{h}$, and equate it to the Gaussian approximation
for $f_{s}$. Solving the equation in the leading logarithmic
approximation yields
$\theta_x\sim\chi_c\sqrt{2\ln\frac{2\chi_c}{\chi'_a}}$, which is of
the order of the scale $\chi_c\sqrt{B}$ at which $f_{h}(\theta_x)$
reaches its maximum. Therefore, around its maximum,
$f_{h}(\theta_x)$ is commensurable with $f_{s}(\theta_x)$, and
consequently, the sum (\ref{fthetax=fhard+fsoft}) needs not develop
a secondary peak or bump. That is what actually happens in practice,
and is physically natural, because a diffusion process tends to
smear out all the features of the probability distribution. (But for
the rescaled distribution $\theta_x^3f(\theta_x)$, as was mentioned
in Sec.~\ref{subsec:Glauber-expansion}, such a bump does exist
\cite{Hanson,Bethe}.)

Figure~\ref{fig:proj-angle} shows the shape of the aggregate
distribution, along with contributions to it from different
mechanisms, for $\chi_c/\chi'_a=10^2$. The figure demonstrates that
the aggregate distribution (solid curve) considerably exceeds the
sum of soft and pure Rutherford components (dot-dashed and dotted
curves, correspondingly). To account for this excess, one has to
employ the resummed hard component (dashed curve) instead of a
single-scattering contribution. So, the issue of resummation of
plural hard scattering contributions is quite essential in practice.
Effectively, it slows down the transition from a Gaussian to
Rutherford regime, so that over a substantial angular interval it
may mimic a law intermediate between Gaussian and Rutherford
decrease, such as a simple exponential law (cf. \cite{Taratin}), or
a power law with an index greater than that for the lowest Born
approximation, as is the case, e.g., for hard scattering of hadrons
(which are themselves composite objects) \cite{Arleo}.


\paragraph{Probabilistic interpretation}

Granted the positivity of both functions $f_{s}(\theta_x)$ and
$f_{h}(\theta_x)$, in conjunction with the normalization condition
$\int^{\infty}_{-\infty} d\theta_x f_{s}+\int^{\infty}_{-\infty}
d\theta_x f_{h}=1$, it is tempting further to interpret them
independently as partial probability distributions. Specifically,
since $f_{h}(\theta_x)$ incorporates all the power-law
contributions, it might be regarded as the probability distribution
of hard-scattered particles, and $f_{s}(\theta_x)$, since it is
nearly Gaussian, should be interpreted as the probability
distribution of soft-scattered particles. That, inevitably, involves
an element of arbitrariness, as long as there is no sharp physical
boundary between soft- and hard-scattered particles. Besides that,
there are regions at sufficiently large $\theta_x$, where
$f_s(\theta_x)$ as evaluated by Eq.~(\ref{fx-soft-def}) becomes
slightly negative [though that is immaterial for practice, because
there it is already overtaken by $f_h(\theta_x)$]. For those
reasons, it is more appropriate to term the encountered functions
\textit{pseudo}-probability distributions. The mentioned
arbitrariness then manifests itself as the residual slight freedom
in the choice of the location of the integration path corner.

\begin{figure}
\includegraphics{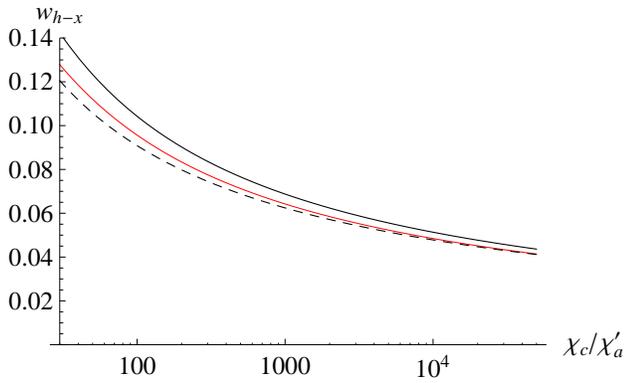}
\caption{\label{fig:wx-hard} Total percentage of hard-scattered
particles in the projected angle distribution, calculated by
Eqs.~(\ref{whard-int}), (\ref{fx-hard-def}), (\ref{approx-saddle-point-eq-proj}) (black solid curve), and by Eqs.~(\ref{whard-int}), (\ref{fx-hard-def}), (\ref{lambda0-x-log-approx}) (red solid curve). Dashed curve, approximation (\ref{whard-log}). }
\end{figure}

Accepting the partial (pseudo-)probability interpretation, let us
assess the corresponding total probability for a particle to belong
to the projected hard component:
\begin{equation}\label{whard-int}
w_{h\text{-}x}(l)=2\int_0^{\infty}d\theta_x
f_{h}(\theta_x,l).
\end{equation}
At large $\chi_c/\chi'_a$, inserting (\ref{fx-hard-def}) to
(\ref{whard-int}) and interchanging the order of integrations leads to
\begin{subequations}
\begin{eqnarray}
w_{h\text{-}x}&=&\frac{2}{\pi\chi_c}\int_0^{\infty}d\nu\sin\frac{\pi\nu^2}4e^{\frac{\nu^2}2\ln\frac{2\chi_c}{\chi'_a\nu}}\nonumber\\
&\,&\qquad\quad\times\int_{\nu\chi_c\ln\frac{2\chi_c}{\nu\chi'_a}}^{\infty}d\theta_x e^{-\frac{\theta_x}{\chi_c}\nu}\nonumber\\
&\underset{\chi_c\gg\chi'_a}\simeq&\frac{1}{2}\int_0^{\sim\chi_c/\chi'_a}d\nu\nu e^{-\frac{\nu^2}2\ln\frac{2\chi_c}{\chi'_a\nu}}.\label{whard-integral}
\end{eqnarray}
The latter single integral can be evaluated by expanding
$e^{\frac{\nu^2}2\ln\frac{\nu}2}\simeq1+\frac{\nu^2}2\ln\frac{\nu}2$,
and integrating termwise, whereupon reassembling it to a single
fraction within the given accuracy:
\begin{equation}\label{whard-log}
w_{h\text{-}x}
\underset{\chi_c\gg\chi'_a}\simeq\frac1{\ln\left(\frac{\chi_c^2}{\chi'^2_a}\ln\frac{\chi_c^2}{\chi'^2_a}\right)-\psi(2)}.
\end{equation}
\end{subequations}
That means that essentially, $w_{h\text{-}x}\simeq1/B$. Formula
(\ref{whard-log}) shows that the fraction of hard-scattered
particles \emph{decreases} with the increase of the target
thickness, as an inverse of its logarithm. The physical reason for
this is that the boundary beginning from which the particles must be
regarded as hard-scattered moves outwards with the increase of the
target thickness, due to the expanding Gaussian component. In
contrast, identity (\ref{int-fk-Moliere=0}) in the Moli\`{e}re
expansion does not grant direct access to the number of particles in
the non-Gaussian component.

The exact behavior of $w_{h}$ as a function of $\chi_c/\chi'_a$ is
plotted in Fig.~\ref{fig:wx-hard} by the solid curve, along with
approximation (\ref{whard-log}) plotted by the dashed curve. It
appears that (\ref{whard-log}) gives a fair approximation for
$w_{h}$ at $\chi_c/\chi'_a\gtrsim10^2$. It may also be mentioned
that the excess of total probability
$w_{s\text{-}x}+w_{h\text{-}x}-1$, for
$w_{s\text{-}x}=2\int_0^{\infty}d\theta_xf_{s}(\theta_x)$ and
$f_{s}(\theta_x)$ evaluated by approximation (\ref{fsoft-x-Gauss}),
(\ref{g0x}), with $C=2.2$, is positive but small compared with
$w_{\text{hard}}$:
\[
w_{s\text{-}x}+w_{h\text{-}x}-1\sim 2\times 10^{-3}.
\]
That corroborates self-consistency of our approximations.

\subsection{Polar angle distribution}

Let us next turn to the somewhat subtler case of the polar angle
distribution, which is given by Bessel integral (\ref{3b}). To appropriately
extend the corresponding diffusion approximation
\begin{equation}\label{f-theta-small-imp-par}
f(\theta,l)\underset{\chi_c/\chi'_a\to\infty}\simeq\frac1{2\pi\chi_c^2}\int_0^{\sim\chi_c/\chi'_a}d\kappa\kappa
J_0\left(\frac{\theta}{\chi_c}\kappa\right)e^{\frac{\kappa^2}2\ln\frac{\chi'_a\kappa}{2\chi_c}}
\end{equation}
to the complex plane of\footnote{In \cite{Bethe}, $\kappa$ was
denoted as $y$, but we keep the same notation as for the projected
angle distribution.} $\kappa=\chi_c\rho$, one needs to substitute
$J_0\left(\frac{\theta}{\chi_c}\kappa\right)=\mathfrak{Re}H_0^{(1)}\left(\frac{\theta}{\chi_c}\kappa\right)$
in the integrand, and exploit the exponential decrease of Hankel
function $H_0^{(1)}(z)$ in the upper half-plane of complex $z$. It
is also preferable in the integrand of (\ref{f-theta-small-imp-par})
\emph{not} to include factor $\kappa$ (physically arising as a part
of the integration element $\kappa d\kappa=d\kappa^2/2$) to the
expression for which the saddle point is sought. Therewith, the
saddle point equation reads
\begin{equation}\label{saddle-point-eq-fullangle}
\frac{\partial}{\partial\kappa}\left[\ln
H_0^{(0)}\left(\frac{\theta}{\chi_c}\kappa\right)+\frac{\kappa^2}2\ln\frac{\chi'_a\kappa}{2\chi_c}
\right]\Bigg|_{\kappa=\kappa_0}=0,
\end{equation}
and like in the previous subsection, its solution at large
$\chi_c/\chi'_a$ must be predominantly imaginary\footnote{That owes
to the fact that $H_0^{(0)}(z)$, like $e^{iz}$, is an even function
of $\mathfrak{Re}z$. This would not be the case if the saddle point
was sought for the integrand including the factor $\kappa$. The
emerging integral representations for $f_{h}$ and $f_{s}$ would then
be too cumbersome.}. Searching a purely imaginary approximation,
i.e., letting $\kappa_0=i\nu_0$, utilizing the relation
$H_0^{(1)}(iz)=\frac{2}{i\pi}K_0(z)$, and neglecting imaginary terms
$\ln i$ compared to the large real logarithm, leads to a real
equation
\begin{equation}\label{approx-saddle-point-eq-fullangle}
\frac{\theta}{\chi_c}\frac{K_1\left(\frac{\theta}{\chi_c}\nu_0\right)}{K_0\left(\frac{\theta}{\chi_c}\nu_0\right)}+\nu_0\left(\ln\frac{\chi'_a\nu_0}{2\chi_c}+\frac12\right)=0.
\end{equation}

\begin{figure}
\includegraphics{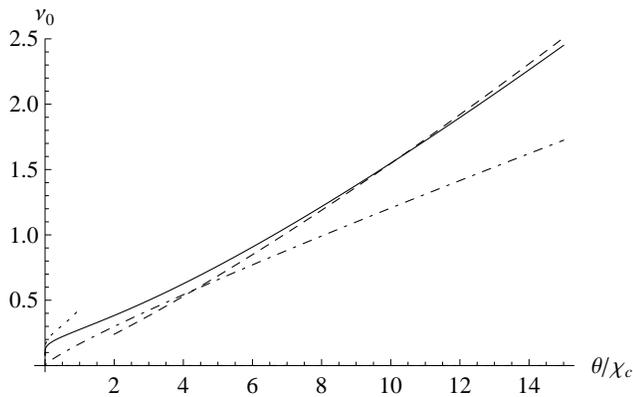}
\caption{\label{fig:lambda0-fullangle} Behavior of the solution of
the corner point equation for the polar angle distribution
[Eq.~(\ref{approx-saddle-point-eq-fullangle})], for
$\chi_c/\chi'_a=10^2$ (solid curve). Dashed curve, approximation
(\ref{lambda0-high-asympt}). Dotted curve, approximation
(\ref{lambda0-low-asympt}). Dot-dashed curve, Bethe's choice for the
corner point, Eq.~(\ref{Bethe-corner-point}).}
\end{figure}

Unfortunately, now Eq.~(\ref{approx-saddle-point-eq-fullangle}) is
difficult to solve by analytic means even approximately, as long as
it requires an approximation for $K_0(z)$ applicable at any positive
$z$. Simple approximations exist only for large $z$, where
$\frac{K_1\left(z\right)}{K_0\left(z\right)}\underset{z\to\infty}\to1$,
implying
\begin{equation}\label{lambda0-high-asympt}
\nu_0\underset{\theta/\chi_c\to\infty}\sim \frac{\theta/\chi_c}{\ln
\left(\frac{2\chi_c^2}{\chi'_a\theta}\ln\frac{2\chi_c^2}{\chi'_a\theta}\right)-1/2}
\end{equation}
[similar to Eq.~(\ref{lambda0-x-log-approx}), and different from
Bethe's choice\footnote{In paper \cite{Bethe}, the saddle point was
actually sought only for part of the integrand,
$K_0\left(\frac{\theta}{\chi_c}\nu\right)e^{\frac{\nu^2}2\ln\frac{2\theta}{\chi'_a
k}}\approx
e^{-\frac{\theta}{\chi_c}\nu+\frac{\nu^2}2\ln\frac{2\theta}{\chi'_a
k}}$ (at real $\nu$, corresponding to purely imaginary $\kappa$).
Eq.~(\ref{Bethe-corner-point}) corresponds to effectively replacing
$\nu$ under the logarithm by $\chi_c/\theta$ rather than
$\theta/\chi_c$, as is suggested by Eq.~(\ref{lambda0-high-asympt}).
That still works when dealing with large-angle asymptotics of the
angular distribution, but not when one aims to find a uniform
approximation for all deflection angles. In the latter case, the
saddle point must be sought for the entire integrand, and the path
corner point be chosen as near as possible to it, as is done in the
present paper. Moreover, even Eq.~(\ref{lambda0-high-asympt}) may be
not the perfect approximation for the entire range of $\theta$ (as
we will see below), so, generally, it seems best to solve the corner
point equation numerically. }
\begin{equation}\label{Bethe-corner-point}
\nu_0=\frac{\theta}{\chi_c\ln\frac{2\theta}{\chi'_a
k}},
\end{equation}
with $k\sim5$], and at small $z$, where
$\frac{K_1\left(z\right)}{K_0\left(z\right)}\sim\frac1{z\ln
\frac1z}$, giving in the leading logarithmic approximation
\begin{equation}\label{lambda0-low-asympt}
\nu_0\underset{\theta/\chi_c\to0}\sim\frac1{\sqrt{\ln\frac{\chi_c}{\theta}\ln\frac{2\chi_c}{\chi'_a}}}.
\end{equation}
The behavior of the solution of
Eq.~(\ref{approx-saddle-point-eq-fullangle}) along with its
asymptotes (\ref{lambda0-high-asympt}), (\ref{lambda0-low-asympt})
is illustrated in Fig.~\ref{fig:lambda0-fullangle}.

\begin{figure}
\includegraphics{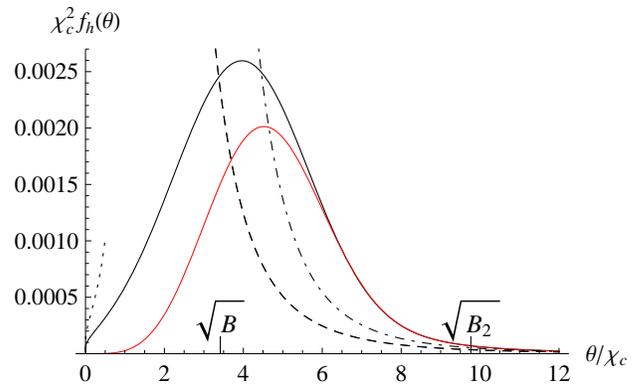}
\caption{\label{fig:f-hard} The shape of the hard scattering
component (\ref{f-hard-def}), (\ref{approx-saddle-point-eq-fullangle}), at $\chi_c/\chi'_a=10^2$ (black solid curve). Red curve, the same for approximate solution (\ref{lambda0-high-asympt}) of the path corner point equation.
Dashed curve, Rutherford asymptotics (\ref{f1-Ruth}); dot-dashed
curve, Rutherford asymptotics with the first power correction,
Eq.~(\ref{R-theta-corr}). Dotted curve, small-angle asymptotics
(\ref{fhard-smalltheta}).}
\end{figure}

Once the solution to Eq.~(\ref{approx-saddle-point-eq-fullangle}) is
found, choosing the integration path similarly to that of
Fig.~\ref{fig:path-proj-angle} leads to a decomposition
\begin{equation}\label{ftheta=fhard+fsoft}
f(\theta,l)=f_{h}(\theta,l)+f_{s}(\theta,l),
\end{equation}
with
\begin{equation}\label{f-hard-def}
f_{h}(\theta,l)=\frac1{\pi^2\chi_c^2}\int_0^{\nu_0(\theta)}d\nu\nu
K_0\left(\frac{\theta}{\chi_c}\nu\right)e^{\frac{\nu^2}2\ln\frac{2\chi_c}{\chi'_a\nu}}\sin\frac{\pi\nu^2}{4},
\end{equation}
and
\begin{equation}\label{f-soft-def}
f_{s}(\theta,l)=\frac1{2\pi\chi_c^2}\mathfrak{Re}\int_{i\nu_0(\theta)}^{\sim\chi_c/\chi'_a}d\kappa\kappa
H_0^{(1)}\left(\frac{\theta}{\chi_c}\kappa\right)e^{\frac{\kappa^2}2\ln\frac{\chi'_a\kappa}{2\chi_c}}.
\end{equation}
Again, we interpret them as partial pseudoprobability distributions
for a particle to belong to hard or to soft scattering probability.
Let us now analyze the behavior of those components, and compare
them with the corresponding projected angle distributions.

\begin{figure}
\includegraphics{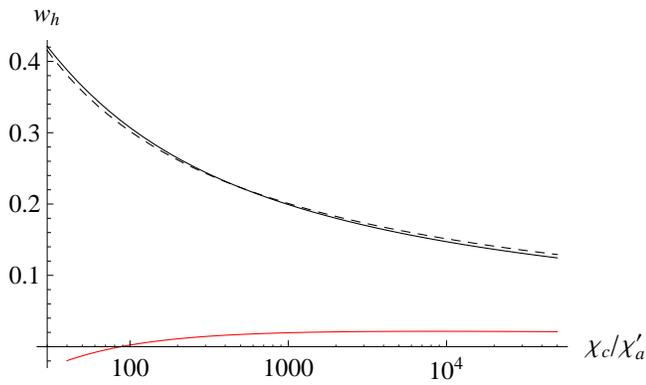}
\caption{\label{fig:w-polar} Total percentage of hard scattered
particles, calculated by Eqs.~(\ref{f-hard-def}),
(\ref{approx-saddle-point-eq-fullangle}) (solid curve). Dashed
curve, interpolation (\ref{wh-heuristic}). The red curve shows the
probability deficit $1-w_{h}-w_{s}$, for $f_{h}(\theta)$ evaluated
by Eqs.~(\ref{f-hard-def}),
(\ref{approx-saddle-point-eq-fullangle}), and $f_{s}(\theta)$ by
Eqs.~(\ref{f-soft-def}), (\ref{fsoft-Gauss-0}) with $C=5$.
 }
\end{figure}

First of all, similarly to the previous subsection, function
$f_{h}(\theta,l)$ proves to be everywhere positive, because typical
$\nu$ for any $\theta$ are less than unity, and then the sine in the
integrand is positive, so we effectively have an integral of a
positive definite function. Using the unlimited growth of $\nu_0$
with $\theta$, it is straightforward to derive the Rutherford
asymptotics for integral (\ref{f-hard-def}), along with its
next-to-leading order power correction:
\begin{eqnarray}\label{R-theta-corr}
f_{h}(\theta,l)&\underset{\theta/\chi_c\to\infty}\simeq&\frac1{4\pi\chi_c^2}\int_0^{\infty}d\nu\nu^3 K_0\left(\frac{\theta}{\chi_c}\nu\right)\nonumber\\
&\,&\qquad\qquad\quad\times\left(1+\frac{\nu^2}2\ln\frac{2\chi_c}{\chi'_a\nu}\right)\nonumber\\
&=&\!\!\frac{\chi_c^2}{\pi\theta^4}+\frac{8\chi_c^4}{\pi\theta^6}\left(\ln\frac{\theta}{\chi'_a}+\gamma_{\text{E}}-\frac32\right).
\end{eqnarray}
The coefficient of the correction term here is in agreement with the
leading log calculation (\ref{f2-asympt}). In the opposite limit
$\theta/\chi_c\to0$, using (\ref{lambda0-low-asympt}), function $f_{h}(\theta)$ can be shown to decrease much slower than
in the case of the projected angle distribution
(\ref{fhard-x-lowthetax}):
\begin{eqnarray}\label{fhard-smalltheta}
f_{h}(\theta,l)&\underset{\theta/\chi_c\to0}\sim&\frac{\nu_0^4}{16\pi\chi_c^2}\ln\frac{\chi_c}{\theta\nu_0}\nonumber\\
&\sim&\frac1{16\pi\chi_c^2
\ln\frac{\chi_c}{\theta}\ln^2\frac{2\chi_c}{\chi'_a}},
\end{eqnarray}
but tends to zero, anyway. Hence, it must reach a maximum at some
finite, nonzero $\theta$. Figure~\ref{fig:f-hard} plots function
(\ref{f-hard-def}) with $\nu_0$ evaluated numerically from
Eq.~(\ref{approx-saddle-point-eq-fullangle}).

In contrast to the case of projected angle distribution, it appears
now that the use of Eq.~(\ref{lambda0-high-asympt}) does \emph{not}
give a good approximation for $f_{h}(\theta)$ simultaneously for all
typical $\theta$ -- because (\ref{lambda0-high-asympt}) is a much
poorer approximation for solution of
Eq.~(\ref{approx-saddle-point-eq-fullangle}) itself. That is
demonstrated by Fig.~\ref{fig:f-hard}, where the red curve
corresponding to approximation (\ref{lambda0-high-asympt}) falls
much below the calculation with the exact solution of
Eq.~(\ref{approx-saddle-point-eq-fullangle}). It signals that for
the polar angle distribution, it is much more reliable to solve the
corner point equation numerically.

Similarly to the previous subsection, we can find that the support
region for function $f_{h}(\theta)$ is concentrated at
\begin{equation}\label{}
\chi_c\sqrt{B}<\theta<\chi_c\sqrt{B_2},\qquad \text{(semihard
region)}
\end{equation}
where $B_2=8B(8e^{2\gamma_{\text{E}}-3}\chi_c^2/\chi'^2_a)$.

\begin{figure}
\includegraphics{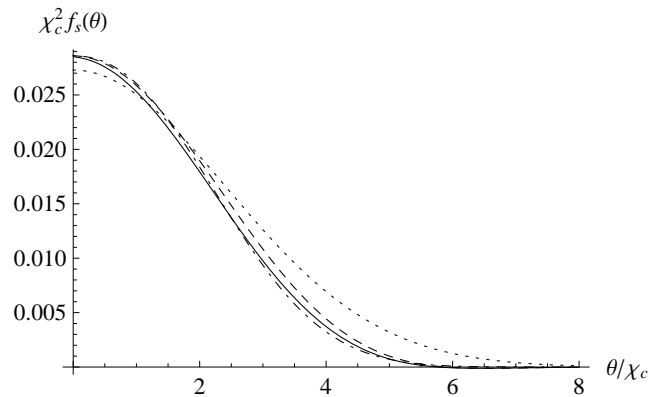}
\caption{\label{fig:soft-polar} The shape of the soft component
(\ref{f-soft-def}) at $\chi_c/\chi'_a=10^2$, built by
Eqs.~(\ref{f-soft-def}) and (\ref{approx-saddle-point-eq-fullangle})
(solid curve). Dashed curve, the same evaluated for the corner point
defined by Eq.~(\ref{lambda0-high-asympt}). Dot-dashed, the
quasi-Gaussian approximation, Eqs.~(\ref{fsoft-Gauss}),
(\ref{fsoft-Gauss-0}), with $C=5$. Dotted, Moli\`{e}re's $f^{(0)}$.
}
\end{figure}

Equation (\ref{approx-saddle-point-eq-fullangle}) also does not
permit expressing $\theta$ through $\nu$, which hampers analytic
computation of the total pseudoprobability of hard scattering
$w_{h}=2\pi\int_0^{\infty}d\theta\theta f_{h}(\theta,l)$ by
interchanging the order of integrations. Numerically, of course,
that presents no difficulty, and is illustrated in
Fig.~\ref{fig:w-polar}. Qualitatively, function
$w_h(\chi_c/\chi'_a)$ exhibits a behavior similar to that of
$w_{h\text{-}x}(\chi_c/\chi'_a)$ in Sec.~\ref{subsec:proj-angle},
but is some 3 times greater, so that it cannot be even regarded as
small. A satisfactory heuristic approximation of the same structure
as Eq.~(\ref{whard-log}) may be written as
\begin{equation}\label{wh-heuristic}
w_{h}\approx\frac{3}{\ln\left(\frac{\chi_c^2}{\chi'^2_a}\ln\frac{\chi_c^2}{\chi'^2_a}\right)-1.5}.
\end{equation}

\begin{figure}
\includegraphics{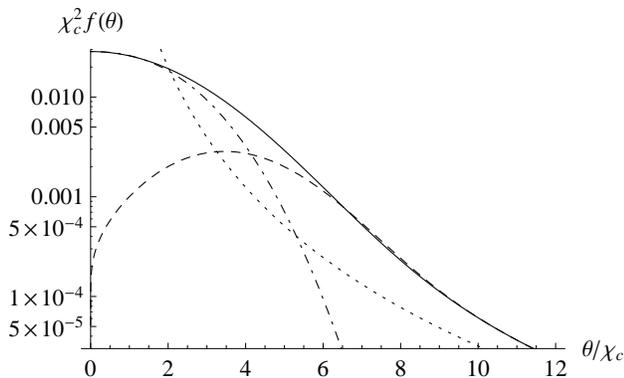}
\caption{\label{fig:full-angle} Relative contributions of the hard
[dashed curve, Eqs.~(\ref{f-hard-def}),
(\ref{approx-saddle-point-eq-fullangle})] and soft [dot-dashed
curve, Eqs.~(\ref{fsoft-Gauss}), (\ref{fsoft-Gauss-0}) with $C=5$]
components to the full-angle distribution [Eq.~(\ref{2DFourier}),
solid curve], for $\chi_c/\chi'_a=10^2$. The sum of thus computed
hard and soft components is virtually indistinguishable from the
solid curve. The dotted curve represents the Rutherford asymptotics
(\ref{f2-asympt}).}
\end{figure}

In what concerns $f_{s}(\theta,l)$, physically it is
expected to exhibit a behavior similar to Eq.~(\ref{fsoft-x-Gauss}).
Indeed, approximation
\begin{equation}\label{fsoft-Gauss}
f_{s}(\theta,l)=f(0,l)e^{-\frac{\theta^2}{2\chi_c^2\ln\left(\frac{2\chi_c^2}{C\chi'_a\theta}\ln
\frac{2\chi_c^2}{\chi'_a\theta}\right)}}
\end{equation}
with
\begin{equation}\label{fsoft-Gauss-0}
f(0,l)\simeq\frac1{2\pi\chi_c^2}\int_0^{\sim\chi_c/\chi'_a}d\kappa\kappa
e^{-\frac{\kappa^2}2\ln\frac{2\chi_c}{\chi'_a\kappa}}
\end{equation}
[an integral similar to (\ref{whard-integral})] and $C\approx5$
works reasonably well (see Figs.~\ref{fig:soft-polar},
\ref{fig:full-angle}). The total pseudoprobability corresponding to
this approximation equals $w_{s}=2\pi\int_0^{\infty} d\theta\theta
f_{s}(\theta)=1-w_{h}-\Delta w$, with $\Delta w\sim 2\times 10^{-2}$
(see Fig.~\ref{fig:w-polar}, red curve), but still tolerably small.
Herein, we will restrict our analysis to this notion.

When comparing the results of this subsection with those of
Sec.~\ref{subsec:proj-angle}, it should be borne in mind
that\footnote{Note that in the lhs and in the rhs of
Eqs.~(\ref{neq}), (\ref{neq-s}), letter $f$ represents different
functions.}
\begin{equation}\label{neq}
f_{h}(\theta_x,l)<\int_{-\infty}^{\infty}d\theta_y
f_{h}(\theta,l)\big|_{\theta=\sqrt{\theta_x^2+\theta_y^2}},
\end{equation}
and correspondingly,
\begin{equation}\label{neq-s}
f_{s}(\theta_x,l)>\int_{-\infty}^{\infty}d\theta_y
f_{s}(\theta,l)\big|_{\theta=\sqrt{\theta_x^2+\theta_y^2}},
\end{equation}
That is clear as long as the integral from a positive function in
the rhs of (\ref{neq}) can not vanish at $\theta_x\to0$, whereas the
left-hand side (lhs) does vanish. It is also natural physically,
because hard collisions which are nearly in the $y$-direction are
not treated as hard when computing the projected distribution in
$\theta_x$. But in the large-$\theta$ asymptotics, (\ref{neq}) holds
as an equality for all the terms of the descending power series, by
virtue of the identity
\begin{equation}\label{}
\frac{2^k
k!}{\pi}\int_{-\infty}^{\infty}\frac{d\theta_y}{\left(\theta_x^2+\theta_y^2\right)^{1+k}}=\frac{(2k-1)!!}{\theta_x^{1+2k}}
\end{equation}
and its derivatives by index $k$, which generate the logarithmic
factors. In turn, inequality (\ref{neq-s}) explains why constant $C$ for approximation (\ref{fsoft-Gauss}) is greater than that for approximation (\ref{fsoft-x-Gauss}).

\section{Summary}

The main conclusions of our paper can be summarized as follows. A
continuation into the complex plane allows presenting the angular
distribution of probability of particles scattered in amorphous
matter as a sum of hard- and soft-scattering components, with no
restriction on the number of scatterings. At that, the hard
component incorporates all the plural-scattering power-law
corrections to the Rutherford single-scattering contribution, while
the soft component is nearly Gaussian, but is narrower than
Moli\`{e}re's $f^{(0)}$. Due to their positivity almost everywhere,
those components admit independent (pseudo)probabilistic
interpretation. The corresponding total percentage of hard-scattered
particles (not appearing naturally in the Moli\`{e}re theory)
amounts typically $w_{h\text{-}x}\sim10\%$ for the projected angle
distribution, and $w_{h}\sim25\%$ in case of the polar angle
distribution), and sets the accuracy limit for Gauss-like
approximations for the soft component.

The second conclusion is that in the aggregate distribution of
scattered particles, there is a significant transition region
between multiple soft and single hard scattering, in which
scattering is multiple but hard. Physically, it is chained to the
fact that at significant target thickness, there always exists a
range of angles, where the probability of several hard rescatterings
is non-negligible. The resummed hard-scattering component peaks at a
non-zero deflection angle, and around its maximum (in the so-called
semihard region), it exceeds the single hard scattering (Rutherford)
contribution by a significant factor. Nonetheless, no bump emerges
in the aggregate distribution around this angle, inasmuch as in the
semihard region, the hard component is comparable with the soft one.

From the practical point of view, it must be noted that if it is
desired to use a single approximation within the central region of
scattering angles only, it may be reasonable to employ Moli\`{e}re's
$f^{(0)}$; but if the hard ``tail" needs description, as well, it is
advantageous to use the separation $f_{h}+f_{s}$ introduced herein.
Even after approximating $f_s$ by a quasi-Gaussian, the sum
$f_{h}+f_{s}$ is still numerically more accurate than a few first
terms of the Moli\`{e}re expansion.

Besides that, it should be remembered that the separation of $f_{h}$
and $f_{s}$ somewhat depends on the choice of the corner point for
the integration path (which does not coincide with the saddle point
of the integrand exactly), and thus may involve slight ambiguity.
Analytic solutions of the corner point equation provide insight into
qualitative dependencies of the particle distribution function on
the total deflection angle and the target thickness, but may
sometimes be insufficiently accurate, so, if better precision is
required, the saddle point equation is to be solved numerically.
Thus, depending on the needs of the study, the proposed construction
may be used either for analytic, or for numerical purposes.





\begin{thebibliography}{00}


\bibitem{Bothe-Wentzel}
W.~Bothe, Z. Phys. \textbf{5}, 63 (1921);

G.~Wentzel, Ann. Phys. (Leipzig) \textbf{69}, 335 (1922).

\bibitem{Williams}

E.J.~Williams, Phys. Rev. \textbf{58}, 292 (1940). 

\bibitem{MoliereFirst}
G.~Moli\`{e}re, Z. Naturforsch. \textbf{2a}, 133 (1947).

\bibitem{Moliere}
G.~Moli\`{e}re, Z. Naturforsch. \textbf{3a}, 78 (1948).

\bibitem{Bethe}
H.A.~Bethe, Phys. Rev. \textbf{89}, 1256 (1953).

\bibitem{Scott}
W.T.~Scott, Rev. Mod. Phys. \textbf{35}, 231 (1963).


\bibitem{Mott-Massey}

N.F. Mott, H.S.W. Massey. \textit{The Theory of Atomic Collisions},
3rd ed. Oxford: Clarendon Press, 1965.


\bibitem{Uch-Zol}
V.V.~Uchaikin, V.M.~Zolotarev. \textit{Chance and Stability. Stable
distributions and their applications} (Utrecht, VSP, 1999).

\bibitem{Bondarenco-paper}

M.V.~Bondarenco, Phys. Rev. D \textbf{90}, 013019 (2014).





\bibitem{Highland-Lynch-Dahl}
V.L.~Highland, Nucl. Instr. Meth. \textbf{129}, 497 (1975).

G.R.~Lynch and O.I.~Dahl, Nucl. Instr. Meth. B \textbf{58}, 6
(1991).

\bibitem{Bond-Shul}

M.V. Bondarenco and N.F. Shul'ga, Phys. Rev. D \textbf{90}, 116007
(2014).


\bibitem{Bielajew}

A.F. Bielajew, Nucl. Instr. Meth. B \textbf{86}, 257 (1994).


\bibitem{Taratin}
A.M.~Taratin and W. Scandale, Nucl. Instr. Meth. B \textbf{335}, 351
(2015).

\bibitem{Arleo}

F. Arleo, S.J. Brodsky, D.S. Hwang, and A.M. Sickles, Phys. Rev.
Lett. \textbf{105}, 062002 (2010).

\bibitem{Hanson}

A.O. Hanson et al., Phys. Rev. \textbf{84}, 634 (1951).

\bibitem{Borel-summ}

See, e.g., M.~Beneke, Phys. Rep. \textbf{317}, 1 (1999).


\bibitem{Fano}
U. Fano, Phys. Rev. \textbf{93}, 117 (1954).
















\end{thebibliography}
\end{document}